\documentclass{ws-procs9x6}

\newcommand{\ltsimeq}{\raisebox{-0.6ex}{$\,\stackrel 
        {\raisebox{-.2ex}{$\textstyle <$}}{\sim}\,$}}

\begin{document}

\title{The Triggering and Bias of Radio Galaxies}

\author{Kate Brand}

\address{National Optical Astronomy Observatory, Tucson, AZ 85726-6732 \\ 
E-mail: brand@noao.edu}

\author{Steve Rawlings}

\address{Astrophysics,Department of Physics, Keble Road, Oxford, OX1 3RH}  

\author{Joe Tufts, Gary J. Hill 
}

\address{McDonald Observatory and Department of Astronomy, University of Texas at Austin, RLM 15.308, Austin, TX 78712}  


\maketitle

\abstracts{We present new results on the clustering and three-dimensional distribution of radio galaxies from the Texas-Oxford NVSS Structure (TONS) survey. The TONS survey was constructed to look at the distribution of radio galaxies in a region of moderate (0 $\ltsimeq$ $z$ $\ltsimeq$ 0.5) redshifts by matching NVSS sources with objects in APM catalogues to obtain a sample of optically bright (R $\le$ 19.5), radio faint (1.4-GHz flux density S$_{1.4} \ge$ 3 mJy) radio galaxies over large areas on the sky. We find that redshift spikes, which represent large concentrations of radio galaxies which trace ($\approx$ 100 $\rm {Mpc}^3$) super-structures are a common phenomena in these surveys. Under the assumption of quasi-linear structure formation theory and a canonical radio galaxy bias, the structures represent $\approx$ 4-5$\sigma$ peaks in the primordial density field and their expected number is low. The most plausible explanation for these low probabilities is an increase in the radio galaxy bias with redshift. To investigate potential mechanisms which have triggered the radio activity in these galaxies - and hence may account for an increase in the bias of this population, we performed imaging studies of the cluster environment of the radio galaxies in super-structure regions. Preliminary results show that these radio galaxies may reside preferentially at the edges of rich clusters. If radio galaxies are preferentially triggered as they fall towards rich clusters then they would effectively adopt the cluster bias.
}

\section{Introduction}

Radio galaxies are ideal probes of large-scale structure as they are biased tracers of the underlying mass and can be easily detected out to high redshifts. By using biased galaxies populations, one can efficiently trace huge super-structures (i.e clusters of clusters of galaxies) which are still in the linear regime and can therefore be directly traced back to rare fluctuations in the initial density field at recombination. However, in order to be useful probes, it is vital to understand how different populations of radio galaxies trace the underlying dark matter (i.e. their bias) and how this has evolved with time. This is directly related to how the radio activity is triggered in different populations and in different environments.

\section{The TONS survey}

The Texas-Oxford NVSS Structure (TONS) survey is a radio galaxy redshift survey comprising three ($\sim$ 25 deg$^2$) independent regions on the sky selected in the same areas as the 7CRS\cite{wil02} and the TexOx-1000 (TOOT) survey\cite{hr}. Unlike 7CRS or TOOT, the TONS survey is selected at 1.4 GHz from the NVSS survey and has fainter radio flux density limits. It also has an optical magnitude limit imposed on it and hence is optimized for looking at clustering of radio galaxies at moderate redshifts ($z \ltsimeq$ 0.5). We obtained optical spectra for all the 84 and 107 radio galaxies in the TONS08 and TONS12 sub-regions respectively. Full details on the the survey selection and observations can be found in \cite{brand}.

\section{Super-structures as traced by radio galaxies}

\begin{figure}
\begin{center}
\setlength{\unitlength}{1mm}
\begin{picture}(150,45)
\put(-2,-10){\includegraphics{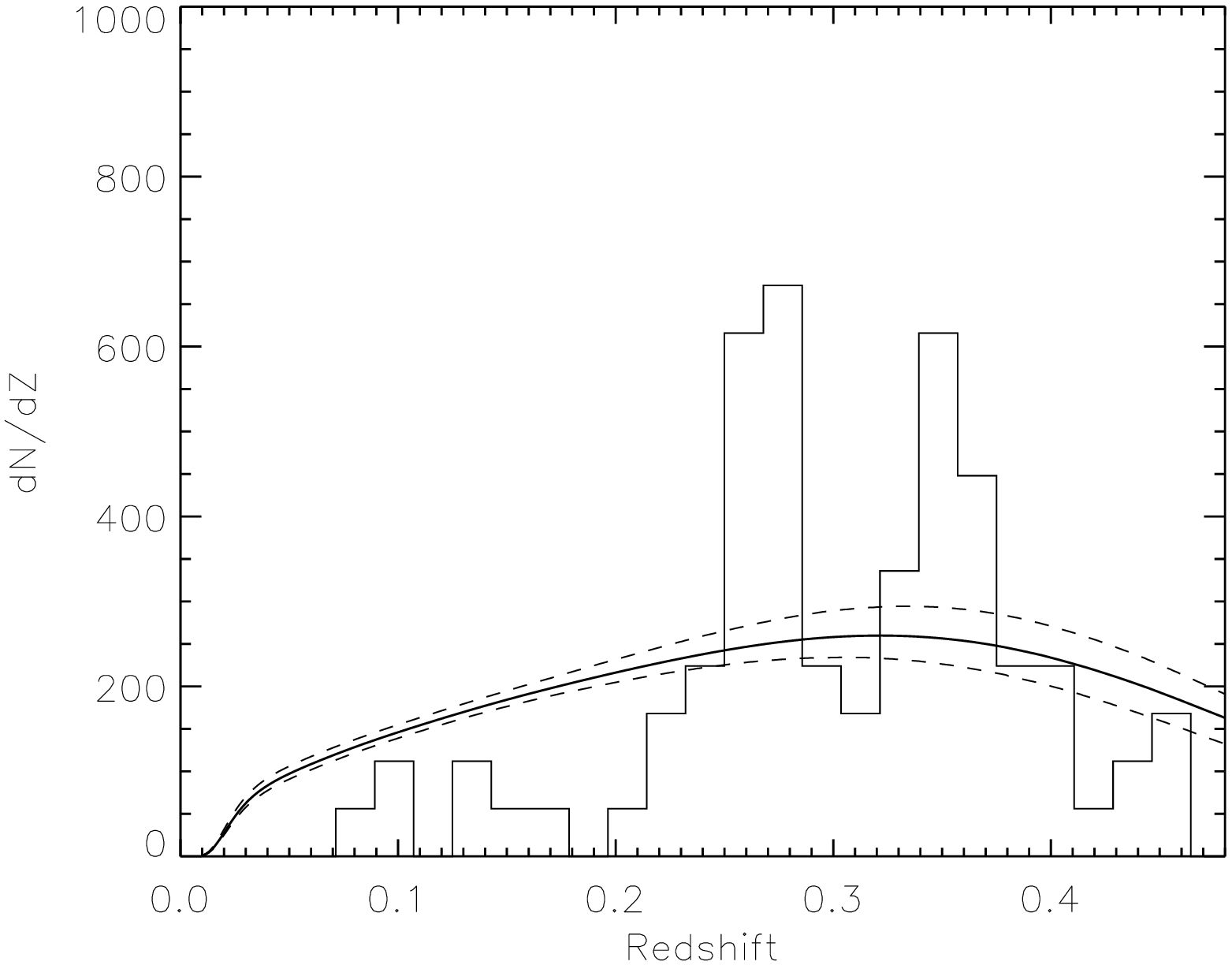}}
\put(55,-10){\includegraphics{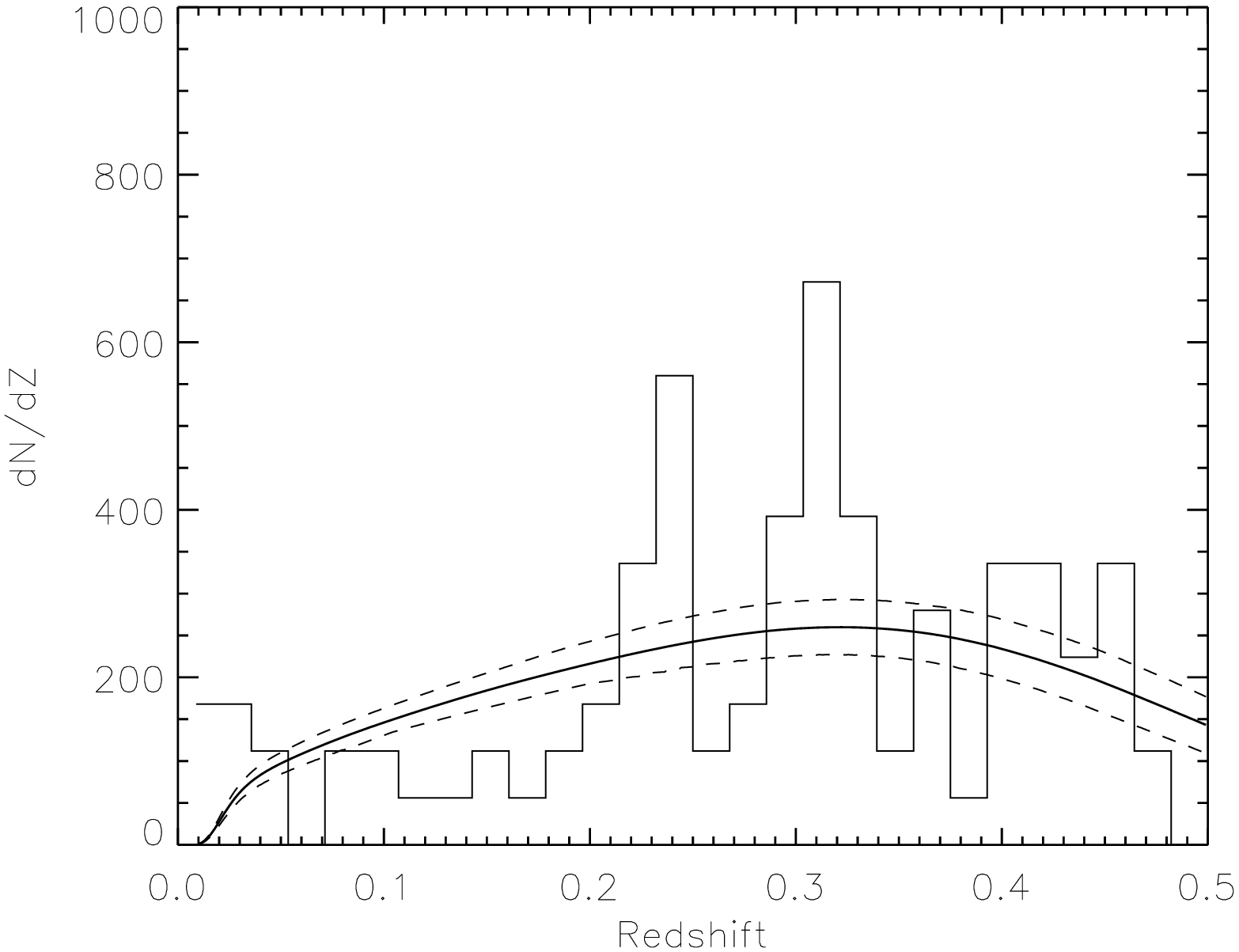}}
\end{picture}
\end{center}
{\caption[junk]{\label{fig:zdist_rlf} The redshift distribution of the TONS08 (left) and TONS12 (right) sub-samples with the model redshift distribution overplotted. The $\pm$1$\sigma$ errors on the model are overplotted (dashed lines). 
}}
\end{figure}

Fig.~\ref{fig:zdist_rlf} shows the redshift distributions of the TONS08 and TONS12 sub-samples. Two significant redshift spikes can been seen at $z\approx$0.27 and $z\approx$0.35 in the TONS08 sub-sample and $z\approx$0.24 and $z\approx$0.32 in the TONS12 sub-sample. It appears that redshift spikes are a common phenomena in this radio galaxy population. These redshift spikes correspond to huge ($\approx$ 100 $\rm {Mpc}^3$) super-structures. 

Assuming the canonical radio galaxy bias\cite{pn}, structure formation theories\cite{bar} predict far fewer super-structures of this size and overdensity than are observed in the TONS survey. The easiest way to reconcile this result is if the radio galaxy bias of this population is larger than the canonical local value. 

\section{The cluster environment of radio galaxies}

Comparing the richness of the environment of the TONS radio galaxies within super-structure regions to those of other radio galaxies should determine whether the large-scale environment is important for triggering of radio activity. For example, the cluster environment may influence the frequency of mergers and interactions between galaxies which may provide fuel to re-ignite the radio emission from the central black hole. Any increase in the triggering of radio activity in dense environments will manifest itself as an increase in the bias of the population. 

$R$-band imaging of all 27 radio galaxies in the $z$=0.27 super-structure in TONS08 survey was performed on the Imaging Grism Instrument (IGI) mounted on the 2.7m Harlan J. Smith telescope at the McDonald Observatory, Texas. We find that the radio galaxies in the TONS08 super-structure are generally in moderately rich (Abell class 0) environments.

\begin{figure}
\begin{center}
\setlength{\unitlength}{1mm}
\begin{picture}(150,45)
\put(-2,-10){\includegraphics{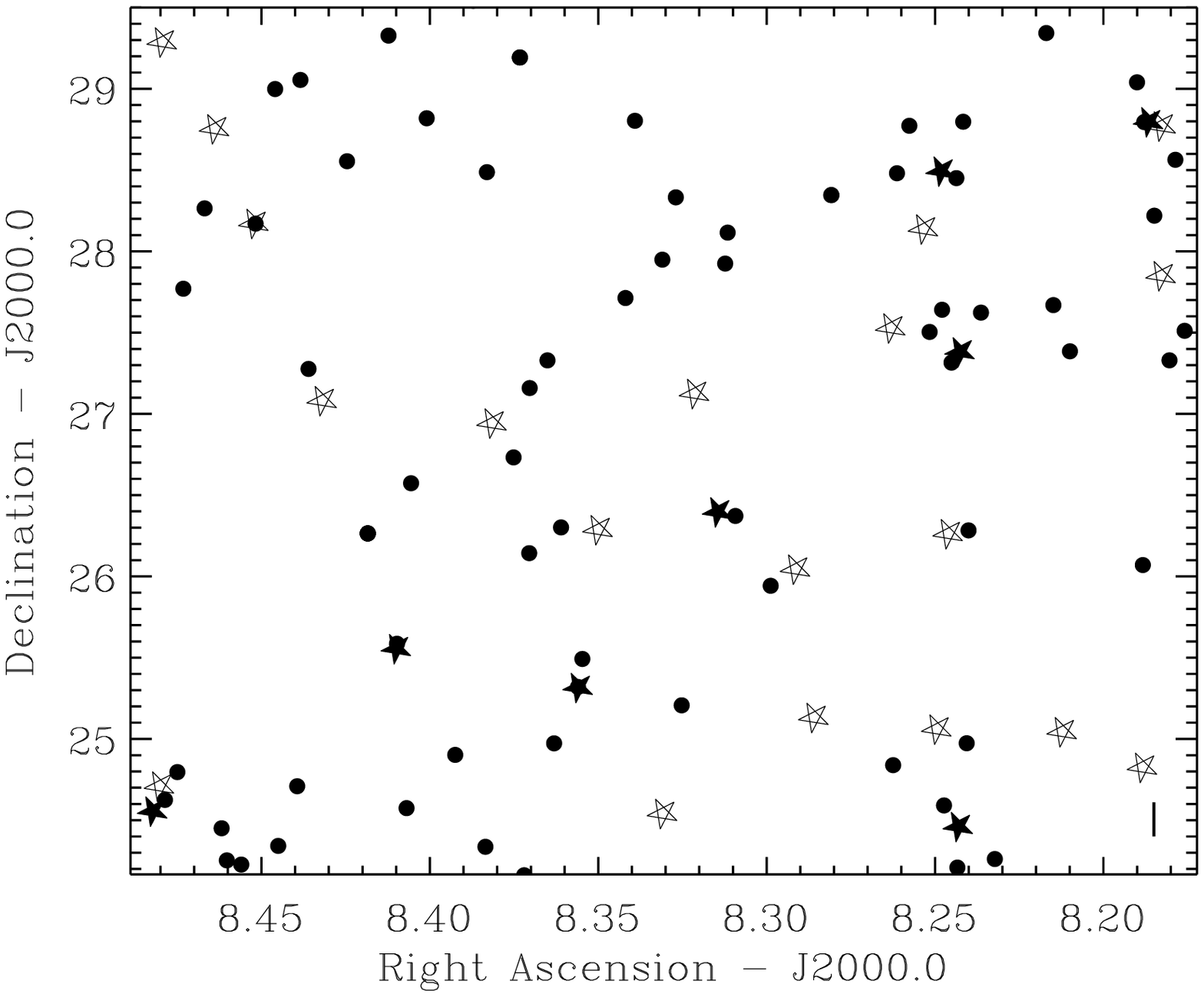}}
\put(55,-10){\includegraphics{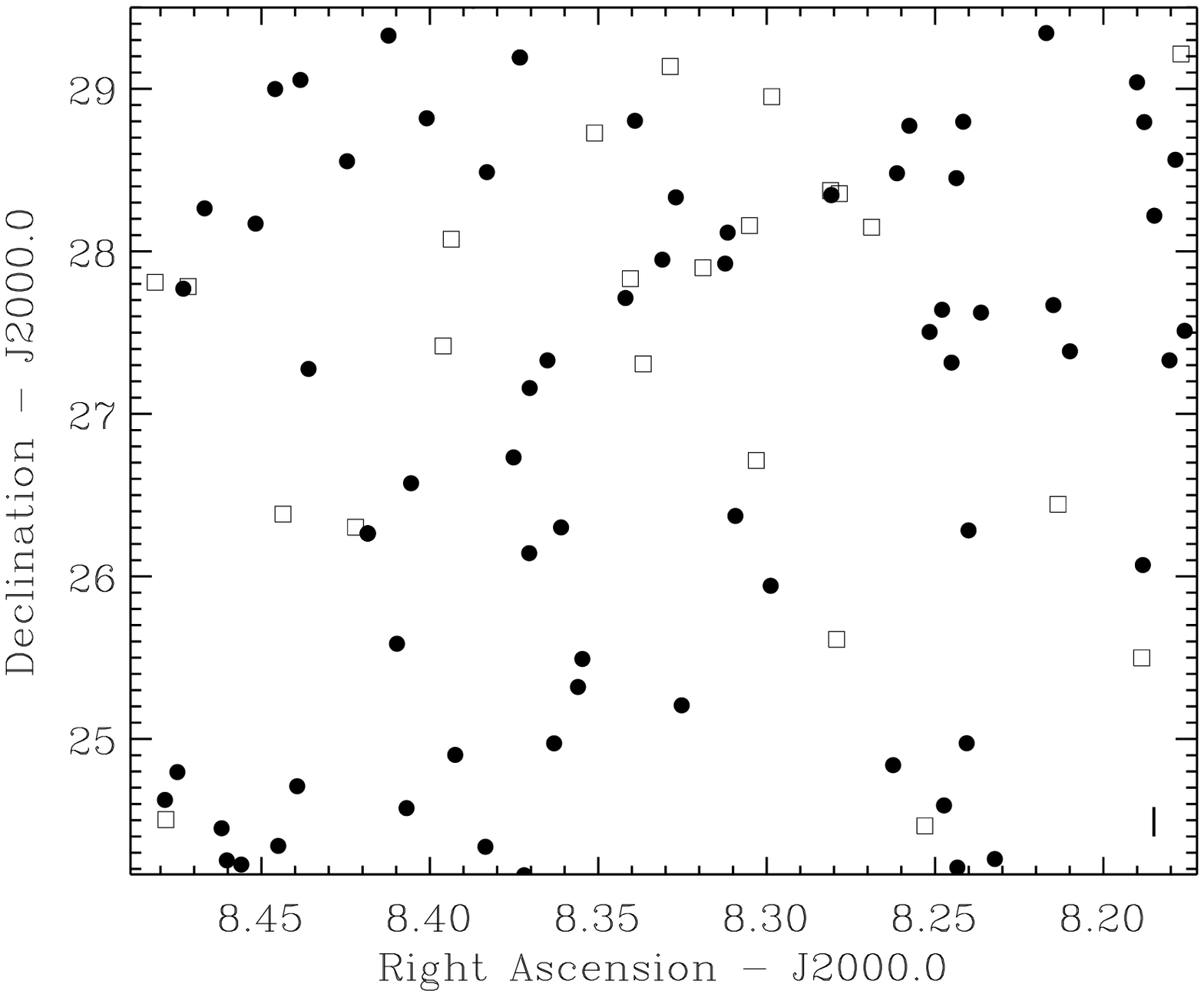}}
\end{picture}
\end{center}
{\caption[junk]{\label{fig:joe} An RA versus DEC plot showing the spatial positions on the sky of the TONS08 $z$=0.27 (left) and $z$=0.35 (right) super-structure radio galaxies with rich (solid stars) and poorer (empty stars) environments. Solid circles show the spatial positions of cluster candidates. The line in the bottom right-hand corner represents the angular size that a length scale of 3 Mpc at $z$=0.27 would have on the sky.}}
\end{figure}

In addition, $R$-band imaging has been performed over the entire TONS08 region using the Prime Focus Corrector (PFC) on the 0.8m telescope at McDonald observatory. Clusters are detected using a matched filter technique\cite{post}. Fig.~\ref{fig:joe} shows the spatial distribution of the cluster candidates and the $z$=0.27 and $z$=0.35 TONS08 super-structure members. All radio galaxies in rich ($B_{\rm{rg}}>$732) environments are within a projected distance of 2.3 Mpc of a cluster candidate. For the $z$=0.27 super-structure, 63 per cent of the radio galaxies are within 3 Mpc (assuming they are at the same redshift) of a cluster candidate. Preliminary results show that in all cases where we have the redshift measurements of cluster candidates near a TONS08 radio galaxy, the redshifts are the same.

\section{Discussion}

The TONS08 radio galaxies within super-structure regions are generally in moderately rich (Abell class 0) environments. However, 63 per cent of the radio galaxies are within a projected distance of 3 Mpc from the centre of a cluster candidate. The fact that we see so many radio galaxies near rich clusters suggests that the radio galaxies are associated with rich clusters but often only on the edges of high overdensity regions. This explains why we find that the radio galaxies are only in moderately rich environments: many of the radio galaxies are further than 0.5 Mpc from the cluster centre. One possible scenario is that of radio galaxies at the centres of poor groups of galaxies being preferentially triggered as the group falls down large-scale structure filaments towards rich clusters. The radio galaxies would then effectively adopt the cluster bias, and the number of redshift spikes we see in the data would become consistent with the number that we expect. 

\hspace{1mm}

\noindent This material is based in part upon work supported by the Texas Advanced Research Program under Grant No. 009658-0710-1999.


\end{document}